\documentclass[aps,pre,showpacs,amsmath,amssymb,amsfonts,lengthcheck,twocolumn,longbibliography,superscriptaddress]{revtex4}
\usepackage{physics} 
\usepackage{bm}
\usepackage{graphicx}
\usepackage{graphics}
\usepackage{amsmath}
\usepackage{amsfonts}
\usepackage{amssymb}
\usepackage{makeidx}
\usepackage[colorlinks=true,citecolor=blue,linkcolor=blue,linktocpage=true,pagebackref=false]{hyperref}
\hypersetup{colorlinks=true,citecolor=blue,linkcolor=blue,filecolor=blue,urlcolor=blue}
\usepackage{color,soul} 
\usepackage{float}
\usepackage{graphicx}
\usepackage{caption}
\usepackage{subcaption}
\usepackage{caption}
\usepackage{graphicx}
\usepackage{ragged2e} 
\usepackage{amsmath}
\usepackage{amssymb}
\usepackage{graphicx}
\usepackage{subcaption}
\usepackage{float} 
\usepackage{xcolor}
\usepackage{tikz}
\usepackage{ragged2e}
 
 \usepackage{orcidlink}

\usetikzlibrary{arrows.meta, calc, positioning, shadows.blur}

\definecolor{hot1}{RGB}{215,67,56}
\definecolor{hot2}{RGB}{140,16,16}
\definecolor{cold1}{RGB}{34,98,189}
\definecolor{cold2}{RGB}{6,44,112}
\definecolor{axisgray}{RGB}{40,40,40}
\definecolor{rojo}{RGB}{255,0,0}

\tikzset{
  >=Latex,
  axis/.style={line width=1.2pt, draw=axisgray},
  cycle/.style={line width=1.1pt, draw=black},
  dashB/.style={draw=black!60, line width=0.9pt, dash pattern=on 2.2pt off 2.2pt},
  flowarrow/.style={-{Latex[length=4mm]}, line width=.11pt, draw=gray},
  heatarrowH/.style={-{Latex[length=3mm]}, line width=1.5pt, draw=hot1},
  heatarrowC/.style={-{Latex[length=3mm]}, line width=1.5pt, draw=cold1},
  pt/.style={circle, inner sep=1.3pt, fill=black},
 resboxH/.style={
    rounded corners=2pt, 
    blur shadow={shadow blur steps=1, shadow xshift=-.1pt, shadow yshift=-0.pt, shadow opacity=0}, 
    inner sep=3.5pt,
    left color=hot2, right color=hot1,shadow scale=0.95, draw=white!70, line width=0.5pt
},
resboxC/.style={
    rounded corners=2pt, 
blur shadow={shadow blur steps=1, shadow xshift=-.1pt, shadow yshift=-0.pt,      shadow scale=0.95, shadow opacity=0.3},
    inner sep=3.5pt,
    left color=cold2, right color=cold1, draw=white!70, line width=0.5pt
}
}

\NewDocumentCommand\eq{ms}{%
  \IfBooleanTF{#2}{\begin{align}#1\end{align}}{\begin{align}
#1
\end{align}}
}
 
\definecolor{rojo}{RGB}{255, 0, 0}

\definecolor{bastian}{RGB}{136,0,123}

\begin{document}
\title{Exact Combinatorial Density of States for the Critical 1D Ising Model}

\author{Bastian Castorene\,\orcidlink{0009-0002-9075-5716}}
\email{bastian.castorene.c@mail.pucv.cl}
\affiliation{Instituto de Física, Pontificia Universidad Católica de Valparaíso, Casilla 4950, 2373223 Valparaíso, Chile}
\affiliation{Departamento de Física, Universidad Técnica Federico Santa María, 2390123 Valparaíso, Chile}
\author{Francisco J. Peña\,\orcidlink{0000-0002-7432-0707}}
\affiliation{Departamento de Física, Universidad Técnica Federico Santa María, 2390123 Valparaíso, Chile}
\author{Martin HvE Groves\,\orcidlink{0009-0000-9711-8467 }} 
\affiliation{Departamento de Física, Universidad Técnica Federico Santa María, 2390123 Valparaíso, Chile}
 \author{Patricio Vargas\,\orcidlink{0000-0001-9235-9747}}
\affiliation{Departamento de Física, Universidad Técnica Federico Santa María, 2390123 Valparaíso, Chile}

\date{\today}

\begin{abstract}
This work presents an exact microcanonical combinatorial analysis of the one-dimensional antiferromagnetic Ising model. At the primary ground-state level crossing $B/J=2$, degeneracies follow the Fibonacci and Lucas sequences for open chains and periodic rings, respectively. We extend this framework to the complete excitation spectrum, demonstrating that the density of states is constructed from topological defects governed by linear Diophantine equations and $p$-fold Fibonacci convolutions. Open boundaries act as fractional defects, densifying the chain spectrum into energy steps of $2J$, whereas the closed ring remains quantized in units of $4J$. Notably, this exact topological counting exposes non-trivial spectral gaps near the fully polarized limit, strictly forbidding the penultimate macroscopic energy levels in both topologies. Through the transfer matrix formalism, we derive exact closed-form expressions for the critical degeneracies at all energy levels. These results provide a rigorous analytical foundation for extracting exact residual entropies and exposing the intrinsic number-theoretic architecture of quantum critical manifolds.
\end{abstract}

\maketitle

\section{Introduction}

\begin{figure*}[t]
\centering
\includegraphics[width=.85\textwidth]{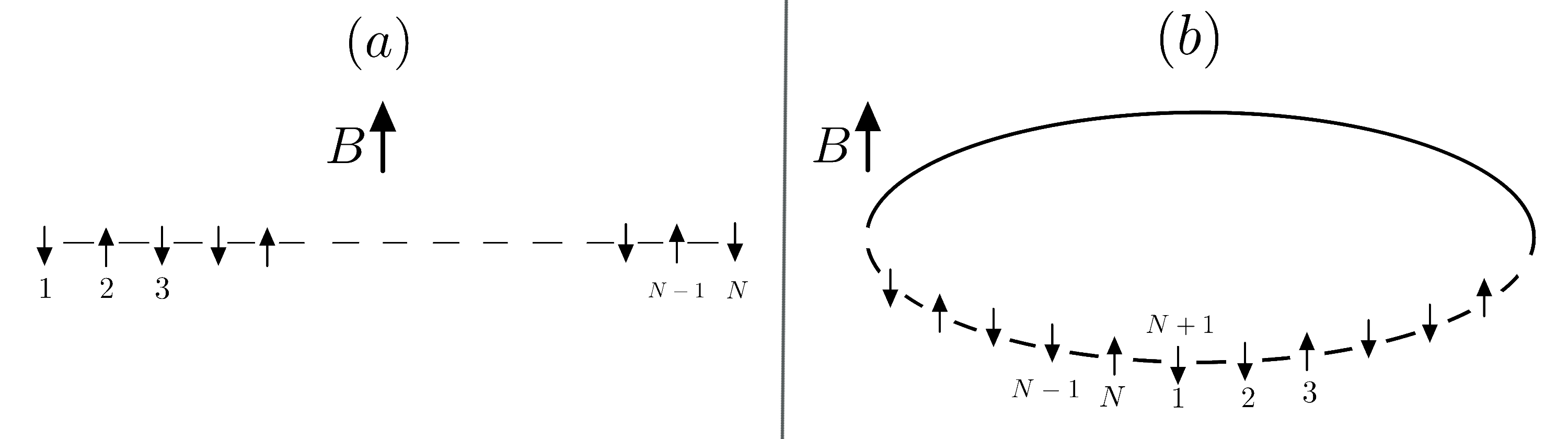}
\caption{(a) Schematic representation of the antiferromagnetic Ising chain with nearest-neighbor interactions under a longitudinal magnetic field $B$ applied along the spin direction. (b) Corresponding ring geometry obtained by imposing periodic boundary conditions.}
\label{fig:ising}
\end{figure*}
   
The Ising model is among the most enduring paradigms in statistical and condensed-matter physics, providing a straightforward yet profoundly rich framework for studying cooperative phenomena, symmetry breaking, and critical behavior in many-body systems~\cite{Brush1967,Baxter1982,Pathria2021,Peliti2011,badalian1996thermodynamics,andrade1997competing,peliti2011statistical}.
Introduced more than a century ago, it has since become a cornerstone of theoretical physics, shaping our understanding of phase transitions, magnetism, and universality classes~\cite{Kadanoff1966,Goldenfeld1992,Binder1987,Sachdev2011}, while simultaneously inspiring developments in areas as diverse as information theory and quantum computation.

Beyond its historical relevance, the Ising model continues to unveil novel conceptual and mathematical connections under modern theoretical perspectives. In particular, the interplay between quantum criticality, topology, and degeneracy has gained renewed interest in quantum thermodynamics~\cite{Vinjanampathy2016,Kosloff2013,Campisi2011,Deffner2019,PhysRevE.93.042111,PhysRevE.56.1371}, as critical manifolds are known to enhance quantum coherence and the efficiency of energy conversion processes. The Ising model has also served as a bridge between quantum and classical descriptions through the Suzuki-Trotter mapping~\cite{Suzuki1976} and has been extended to competing-interaction and frustrated systems exhibiting multiparametric quantum criticality~\cite{Ovchinnikov2003,Dutta2015,Kassan-ogly01122001,Wang_2019,Corona_2016,nachtergaele1988ground}. In these contexts, the emergence of Fibonacci and Lucas-type relations frequently signals underlying recursive or topological symmetries, reflecting the deep link between discrete mathematics and critical phenomena~\cite{Baez2011,Baxter1982,redner}. 

In its one-dimensional form, the Ising model remains analytically tractable through integral methods within the canonical ensemble, and its properties continue to be explored using various analytical techniques~\cite{PhysRevB.27.4503,Hauspurg_2024,huang2008statistical}. Historically, Kramers and Wannier demonstrated that the partition function of a 2D-lattice model could be expressed as the trace of a product of transfer matrices~\cite{KramersWannier1941}. Building upon this framework, the two-dimensional Ising model, which had previously resisted a closed-form analytical solution, was exactly solved by Onsager~\cite{Onsager1944} by determining the eigenvalues of the associated transfer matrix to derive the exact thermodynamic quantities. Subsequently, the one-dimensional model has been revisited in multiple modern contexts, ranging from renormalization and scaling theory~\cite{Goldenfeld1992,Suzuki1976} to quantum information and fidelity approaches to phase transitions~\cite{Gu2010,Zanardi2006,Quan2006}.

Experimentally, Zhang \textit{et al.} demonstrated the direct observation of quantum criticality in finite Ising spin chains using nuclear magnetic resonance (NMR) quantum simulators~\cite{PhysRevA.79.012305}, revealing the sequence of level crossings and critical fields associated with the antiferromagnetic-to-paramagnetic transition. From a theoretical standpoint, Silva da Conceição and Maia~\cite{PhysRevE.96.032121} established a formal connection between the one-dimensional Ising partition function and generalized Lucas polynomials, showing that the canonical partition function obeys a recurrence relation of the Fibonacci-Lucas type. Together, these contributions highlight both the experimental accessibility of critical degeneracies and their algebraic support within the canonical ensemble.

However, an exact analytical micro-canonical distribution of the density of states for the antiferromagnetic Ising model throughout the entire critical excitation spectrum has yet to be established. Because this methodology is inherently rooted in the enumeration of micro-states, analytic combinatorics plays a fundamental role in its mathematical formulation. The primary tools rely on symbolic methods and combinatorial transfer theorems; while these differ conceptually from the physical transfer-matrix derivation, they are ultimately isomorphic \cite{flajolet2009analytic}. The formal construction of sequences, compositions, and multi-sets links microscopic configurations to exact combinatorial coefficients. For instance, the number of compositions of an integer $n$ into exactly $c$ positive parts introduces nontrivial mathematical relations intricately connected to $p$-fold Fibonacci convolutions~\cite{Stanley2012,Comtet1974}.

Consequently, diverse challenges---ranging from the non-adjacency counting problem in linear arrangements to enumerating the independent vertex subsets of a graph---are fundamentally related to the Fibonacci and Lucas sequences~\cite{BenjaminQuinn2003,ProdingerTichy1982,Koshy2001,Bunder01051992}. Furthermore, the requirement that discrete configurations satisfy structural equivalences can be directly modeled via linear Diophantine equations~\cite{niven1991introduction}.

In this context, understanding the exact combinatorial structure of degenerate states sharing the same zero-temperature energy provides a fundamental bridge between microscopic spin configurations and macroscopic thermodynamic quantities, such as residual entropy~\cite{Baez2011,Baxter1982,Pathria2021,Lieb1967,Pauling1935}. Furthermore, establishing the density of states analytically offers a profound advantage over numerical sampling methods. While standard Metropolis Monte Carlo simulations sample individual configurations for a given energy, flat-histogram techniques like the Wang-Landau algorithm~\cite{10.1119/1.1707017} reconstruct the energy landscape by computing the macroscopic degeneracies. Deriving these degeneracies exactly provides a thermodynamically richer description, directly yielding the exact total entropy of the system and bypassing numerical approximations entirely \cite{WangLandau1}.

Diverging from approaches focused on the canonical partition function or the dynamical detection of criticality, we perform a microcanonical combinatorial analysis of the one-dimensional antiferromagnetic Ising model under a longitudinal magnetic field. We show that at the nontrivial ground-state level crossing (GLC) $B_{\mathrm{crit}} = B/J = 2$, the number of degenerate configurations follows the Fibonacci sequence for open chains and the Lucas sequence for periodic rings. Crucially, we extend this analysis to the full excitation spectrum, where a Diophantine equation dictates how spin pairs organize across the various excited levels, exposing a fundamental structural distinction: open boundaries act as fractional topological defects that densify the chain spectrum into energy steps of $2J$, whereas the closed-ring spectrum remains quantized in discrete units of $4J$.
To systematize the state counting, we apply tools from analytic combinatorics, demonstrating that the critical manifold is constructed through a superposition of boundary collapses and bulk clusters governed by $p$-fold Fibonacci convolutions~\cite{Fibo1,Alexander2014AGE,Bunder01051992,flajolet2009analytic}. By mapping these topological defects onto an algebraic generating function via the transfer matrix formalism, we establish an exact generalization for the density of states across all excitation levels at criticality. In contrast to prior algebraic treatments, our analysis highlights the number-theoretic structure of critical degeneracies as an intrinsic physical property of the Ising Hamiltonian, suggesting broader implications for one-dimensional spin chains, critical manifolds, and the combinatorial properties of quantum many-body systems.



\section{Model} 
The working substance consists of $N$ spin particles described by a one-dimensional antiferromagnetic Ising model subjected to a longitudinal magnetic field. The Hamiltonian governing the system is given by
\begin{align}
\hat{\mathcal{H}} &= J \sum_{i=1}^{N-1} \sigma_i^z \sigma_{i+1}^z + B \sum_{i=1}^{N} \sigma_i^z , \label{H}
\end{align}
where $J$ denotes the antiferromagnetic exchange coupling between nearest-neighbor spins $(J > 0)$ and $\sigma_i^z$ represents the Pauli operator along the $z$ direction acting on the $i$th spin, whose eigenvalues are $\sigma_i = \pm 1$. As illustrated in Fig.~\ref{fig:ising}(a), the system forms a linear chain of interacting spins. In this configuration, both translational symmetry $\sigma_i \leftrightarrow \sigma_{i+\delta}$ and spin-reflection symmetry $\sigma_i \rightleftharpoons -\sigma_i$ are simultaneously broken \cite{peliti2011statistical}. The longitudinal magnetic field $B$ is applied along the positive $z$ axis and, in physical units, is expressed as $B = \mu_B g_z H$, where $\mu_B$ is the Bohr magneton and $g_z$ the Landé $g$-factor. For notational simplicity, all field values are expressed in units of the exchange constant throughout this work.

By extending the summation from $(N-1)\rightarrow N$ in the exchange term and imposing periodic boundary conditions, $\sigma^z_{N+1}=\sigma^z_1$, the open-chain topology is transformed into a closed ring, as shown in Fig.~\ref{fig:ising}(b), while the magnetic-field contribution remains unchanged. 
\begin{figure}
    \centering
\begin{tikzpicture}[>=Stealth, line cap=round, scale=0.9,every node/.style={scale=1.1}]
  \draw[->,line width=2pt] (0,0) -- (0,6.2) node[above] {\Large $E$};r
  \draw[->,line width=2pt] (0,0) -- (8.2,0) node[right] {\Large $B$};

  \coordinate (GLC) at (1,4.5);

  \def\mblue{-0.11}    
  \def\morange{-0.5}   

  \def\mgrayA{ 0.1}   
  \def\mgrayB{ 0.05}   
  \def\mgrayC{-0.01}   

  \def\xend{8}


  \def\xcross{4.2}
  \def\ycross{5.0}
  
  \draw[gray!50, thin] (1,{\ycross + \mgrayA*(1-\xcross)}) -- (8,{\ycross + \mgrayA*(8-\xcross)});
  \draw[gray!50, thin] (1,{\ycross + \mgrayB*(1-\xcross)}) -- (8,{\ycross + \mgrayB*(8-\xcross)});
  \draw[gray!50, thin] (1,{\ycross + \mgrayC*(1-\xcross)}) -- (8,{\ycross + \mgrayC*(8-\xcross)});

  \def\xcross{4.2}
  \def\ycross{4.0}
  
  \draw[gray!50, thin] (1,{\ycross + \mgrayA*(1-\xcross)}) -- (8,{\ycross + \mgrayA*(8-\xcross)});
  \draw[gray!50, thin] (1,{\ycross + \mgrayB*(1-\xcross)}) -- (8,{\ycross + \mgrayB*(8-\xcross)});
  \draw[gray!50, thin] (1,{\ycross + \mgrayC*(1-\xcross)}) -- (8,{\ycross + \mgrayC*(8-\xcross)});

  \def\mgrayD{-0.1}
  \def\mgrayE{-0.05}
  \def\mgrayF{0.01}
  
  \def\xcross{4.2}
  \def\ycross{4.4}

  \draw[gray!50, thin] (1,{\ycross + \mgrayD*(1-\xcross)}) -- (8,{\ycross + \mgrayD*(8-\xcross)});
  \draw[gray!50, thin] (1,{\ycross + \mgrayE*(1-\xcross)}) -- (8,{\ycross + \mgrayE*(8-\xcross)});
  \draw[gray!50, thin] (1,{\ycross + \mgrayF*(1-\xcross)}) -- (8,{\ycross + \mgrayF*(8-\xcross)});
  \draw[line width=1.2pt,blue] (1,2.5) -- (8,{4.5 + \mblue*(8+18)});
  \draw[line width=1.2pt,white!10!white,dashed,opacity=0.5] (1,4.5) -- (8,{4.5 + \mblue*(8-1)});

  \draw[line width=1.2pt,orange] (1,3.5) -- (8,{4.5 + \morange*(8+.1)});
  \draw[line width=1.2pt,red!70!black,dashed,opacity=0.6] (1,3.5) -- (8,{4.5 + \morange*(8+.1)});
  
  
    \draw[line width=1.2pt,gray] (1,3.) -- (8,{4.5 + \morange*(8-1.1)});
   \draw[line width=1.2pt,black!70!black,dashed,opacity=0.6] (1,3.) -- (8,{4.5 + \morange*(8-1.1)});



  \filldraw[black] (4.2,2.1) circle (2.8pt);
  \node[anchor=west,yshift=13pt,xshift=3pt] at (4.2,2.0) {LC};

  \draw[dashed] (1,0) -- (1,5.6);
  \node[below] at (4.2,0) {\Large $B_{\mathrm{crit}}$};

  \draw[dotted,line width=1.0pt] (4.2,0) -- (4.2,{7.8 + \morange*(5.6-1)});
   \node[below] at (1,0) {\Large $0$};

\end{tikzpicture}
\caption{Illustrative energy-level diagram of the Ising model as a function of the magnetic field $B$. An LC is observed at $B_{\text{crit}}=B/J=2$, where multiple energy levels become degenerate.}
    \label{fig:energias}
\end{figure}
Fig.~\ref{fig:energias} illustrates the distribution of the $2^N$ energy levels of the one-dimensional Ising model as a function of the magnetic field. Depending on the parity of $N$, the Ising chain exhibits three ground-state level crossings (GLCs) for even system sizes and two for odd system sizes~\cite{PhysRevB.68.214406}. Specifically, an additional parity-dependent critical point emerges at $B/J=1$ (see Appendix~\ref{parity}). In all cases, however, a nontrivial LC occurs at $B_{\text{crit}} = B/J = 2$, where multiple energy levels cross for $N\geq 2 $. At this LC, the system undergoes a transition from an antiferromagnetic to a paramagnetic phase as the spins align with the external magnetic field, leading to a nontrivial degeneracy involving several energetically equivalent configurations.

In particular, the energy contributions of these degenerate states are primarily determined by the relative orientation of the spins composing the chain (ring). From Fig.~\ref{fig:energias}, one can infer that several spin configurations ${\sigma_i}$ share the same total energy, as schematically illustrated in Fig.~\ref{fig:ising}. The distinct energy configurations depend solely on the number of up and down spins, $N_\uparrow$ and $N_\downarrow$, the number of parallel nearest-neighbor pairs $N_{\uparrow\uparrow}$ and $N_{\downarrow\downarrow}$, and the number of domain walls $N_{\uparrow\downarrow}$ \cite{huang2008statistical}. These quantities follow  $
N_\uparrow + N_\downarrow = N$.
Defining the coordination number $q=2$ for both topologies \cite{pathria}, the $2N_\uparrow$ total bonds originating from up spins must either connect to another up spin (sharing two bonds per $\uparrow\uparrow$ pair) or to a down spin (one bond per $\uparrow\downarrow$ pair). By symmetry, the same logic applies to down spins, yielding:
\begin{align}
\begin{aligned}
    2N_\uparrow &= 2N_{\uparrow\uparrow} + N_{\uparrow\downarrow},    \\
2N_\downarrow &= 2N_{\downarrow\downarrow} + N_{\uparrow\downarrow} .
\end{aligned} \label{spinupup}
\end{align}
Thus, the total number of spins can be written in terms of nearest-neighbor combinations:
\begin{align}
N_{\uparrow\uparrow} + N_{\downarrow\downarrow} + N_{\uparrow\downarrow} = N .
\end{align}

Substituting these relations into the Ising Hamiltonian Eq.~(\ref{H}) yields an equivalent representation in terms of spin populations:
\begin{align}
\mathcal{H} = J\left(N_{\uparrow\uparrow} + N_{\downarrow\downarrow} - N_{\uparrow\downarrow}\right) + B\left(N_\uparrow - N_\downarrow\right) . \label{H_N}
\end{align}

Aligned spins contribute $+1$ and anti-aligned spins $-1$ to the exchange interaction energy. Similarly, the magnetic-field term minimizes (maximizes) the energy when the spins point downward (upward), contributing $-1$ and $+1$, respectively. Expressing the Hamiltonian Eq.~(\ref{H_N}) in terms of the number of up spins, domain walls, and total sites yields
\begin{align}
\mathcal{H} = J\qty(N - 2N_{\uparrow\downarrow}) + B\qty(2N_\uparrow - N) . \label{H_enterminosdeN}
\end{align}

    Focusing on the ground state at the LC  with $B_{\text{crit}} = B/J = 2$ in Fig.~\ref{fig:energias}, the minimum energies of the open chain and the closed ring are given by
\begin{align}
\begin{aligned}
    \mathcal{H}^{\text{chain}}   &= -(N+1) + 4N_\uparrow - 2N_{\uparrow\downarrow}   , \\
\mathcal{H}^{\text{ring}}   &= -N + 4N_\uparrow - 2N_{\uparrow\downarrow} . 
\end{aligned}\label{N_combinaciones1}
\end{align}

A particular case of minimum energy arises when no spins point up, and hence no domain walls are present:
\begin{align}
\begin{aligned}
    \mathcal{H}_{\text{ground}}^{\text{chain}} &= -(N+1) , \\
\mathcal{H}_{\text{ground}}^{\text{ring}}  &= -N .
\end{aligned} \label{Ground_Energies}
\end{align}

Consequently, the open-chain configuration exhibits a slightly lower ground-state energy than the closed ring for the same number of sites $N$. The next step is to determine the number of distinct microstates leading to the same ground-state energy, i.e., to find the number of distinct ways in which $N$ spins in the chain or ring can be arranged to yield the same ground energy of Eq.~\eqref{Ground_Energies}.

\section{Fibonacci-Lucas Ground Degeneracies}

To determine the number of possible microstates corresponding to spin-up configurations in the lowest-energy state, the ground-state energies in Eqs.\eqref{Ground_Energies} are equated to the spin-based Hamiltonian in Eqs.\eqref{N_combinaciones1}. This leads to the same condition for both the open chain and the closed ring:
\begin{align}
2N_{\uparrow} = N_{\uparrow\downarrow} \implies N_{\uparrow\uparrow} = 0 , \label{condicion_spinup}
\end{align}
which indicates that no two spin-up sites can be adjacent when this condition is imposed on Eqs.~\eqref{spinupup} implies that spin-up sites must be separated by at least one spin-down site. 

For the open chain, the number of admissible up-spins ranges from zero up to the integer part of $(N-1)/2$, strictly excluding the boundary sites to maintain the condition in Eq.~\eqref{condicion_spinup}. Since the edge spins have only a single neighbor, an up-spin at either boundary can form only one up-down bond, contributing a single domain wall. To satisfy the constraint $2N_{\uparrow} = N_{\uparrow\downarrow}$, each up-spin must generate exactly two domain walls, which physically prohibits up-spin occupation at the boundaries. Consequently, the valid configurations correspond to the number of ways to arrange the spins so that no two up-sites are adjacent, which is structurally analogous to the combinatorial problem of distributing non-consecutive elements on a discrete linear chain. The resulting combinatorial expression in the microcanonical ensemble is

\begin{align}
\Omega_{\text{ground}}^{\text{chain}} = \sum_{N_\uparrow=0}^{\left\lfloor\frac{N-1}{2}\right\rfloor} \binom{N-1-N_\uparrow}{N_\uparrow}
=  F_N  . \label{omega_fibo1}
\end{align}
where the sum in Eq.~\eqref{omega_fibo1} defines the $N$-th Fibonacci number ($F_N$) \cite{Fibo1}.

In the case of the closed ring, applying the same non-adjacency condition given in Eq.~\eqref{condicion_spinup} connects the boundary sites, introducing circular symmetry. This symmetry allows additional spin-up placements that are forbidden in the open chain. The number of admissible configurations, ranging from zero up to the integer part of half the system size $N/2$, is therefore analogous to counting the number of ways to arrange heads on a closed ring of $N$ coins such that no two heads are adjacent:
\begin{align}
\Omega_{\text{ground}}^{\text{ring}} = \sum_{N_\uparrow=0}^{\left\lfloor\frac{N}{2}\right\rfloor}
\frac{N}{N-N_\uparrow} \binom{N-N_\uparrow}{N_\uparrow}
 = L_N\label{omega_lucas1}
\end{align}
The combinatorial factor $  {N}/{(N-N_\uparrow)}$ accounts for the periodic boundary and rotational symmetry of the ring. By definition, this expression represents the $N$-th Lucas number ($L_N$) \cite{Alexander2014AGE}.

The number of ground-state microstates in the ring, given by Eq.\eqref{omega_lucas1}, exceeds that of the open chain, described by Eq.\eqref{omega_fibo1}.

\section{Higher-Order Excitations} 
\subsection{Combinatorial Structure of the Excited States}

The procedure to determine the exact combinatorial structure of the excited states relies on evaluating the discrete energy gap generated by perturbing the topological constraints of the ground state. At the Level Crossing (LC) with $B_{\mathrm{crit}} = 2$, the energy of any arbitrary configuration relative to the absolute minimum can be analytically deduced.

For the open chain, the total number of up spins decomposes into interior ($N_{\uparrow,\text{int}}$) and boundary ($N_{\uparrow,b}$) contributions, such that
\begin{equation}
    N_\uparrow = N_{\uparrow,\text{int}} + N_{\uparrow,b}.
\end{equation}
The total number of domain walls is related to the internal and boundary up spins. Inserting this decomposition into Eq.~\eqref{spinupup}, we obtain
\begin{equation}
    N_{\uparrow\downarrow} = 2N_{\uparrow,\text{int}} + N_{\uparrow,b} - 2N_{\uparrow\uparrow}.
\end{equation}
Substituting these relations into the Hamiltonian in Eq.~\eqref{N_combinaciones1} evaluated at $B/J=2$, the energy shifts relative to the ground state in Eq.~\eqref{Ground_Energies} are given by
\begin{align}
    \Delta E^{\text{chain}} &= J(2N_{\uparrow, b} + 4N_{\uparrow\uparrow}), \label{gap_chain} \\
    \Delta E^{\text{ring}} &= 4J N_{\uparrow\uparrow}. \label{gap_ring}
\end{align}

Eq.~\eqref{gap_chain} dictates that the open-chain ground state ($\Delta E^{\text{chain}} = 0$) is strictly achieved when both boundary spins are pointing down ($N_{\uparrow, b} = 0$) and no adjacent up spins exist ($N_{\uparrow\uparrow} = 0$), recovering the results from Eq.~\eqref{omega_fibo1}. For the closed ring, periodic boundary conditions naturally forbid boundary spins, rendering the energy gap dependent exclusively on the number of bulk adjacent pairs, as shown in Eq.~\eqref{gap_ring}.

\textit{Spectral Pattern and Change of Variables:}
The energy gap equations directly reveal the quantization of the excitation spectrum through the available topological defects. To systematize the combinatorial analysis, we introduce a change of variables, defining the topological indices $b \equiv N_{\uparrow, b}$ and $k \equiv N_{\uparrow\uparrow}$. For the open chain, the discrete excitation index $m$ is determined by the linear Diophantine equation \cite{niven1991introduction}:
\begin{equation}
    m = b + 2k. \label{m_restriction}
\end{equation}
Since the number of excited boundaries is strictly bounded as $b \in \{0, 1, 2\}$ and the number of bulk adjacent pairs is $k \in \mathbb{N}_0$, the integer solutions to this Diophantine equation classify all permissible macroscopic configurations for each excitation level $m \in \mathbb{N}_0$. Conversely, for the closed ring, the excitation index is governed solely by bulk defects, defined directly as $k \in \mathbb{N}_0$.

Expressing the energy gaps in terms of these new indices, the excitation spectrum of the one-dimensional Ising model at the LC $B/J=2$ rigorously follows:
\begin{align}
    \Delta E_m^{\text{chain}} &= 2mJ, \quad m \in \mathbb{N}_0, \label{pattern_chain} \\
    \Delta E_k^{\text{ring}} &= 4kJ, \quad k \in \mathbb{N}_0. \label{pattern_ring}
\end{align}

Eqs.~\eqref{pattern_chain} and \eqref{pattern_ring} demonstrate that the closed ring spectrum is rigorously quantized in discrete units of $4J$, representing the minimum energy cost to violate the bulk non-adjacency constraint. The open chain, however, exhibits a strictly denser spectrum quantized in half-steps of $2J$. This discrepancy indicates that the open boundaries function as fractional topological defects; a boundary spin flip costs exactly half the energy of an adjacent bulk pair, permanently altering the macroscopic state counting and the thermodynamic limit of the critical manifold.

Furthermore, the Diophantine relation in Eq.~\eqref{m_restriction} intrinsically bounds the excitation spectrum. For an open chain of length $N$, the finite availability of boundary and bulk defects strictly restricts the system to exactly $2N-1$ distinct energy levels, bounding the excitation index to $m \leq 2N-2$ for the macroscopic degeneracies.

\subsection{First-excited state}
 
The degeneracy of this first excited state, $\Omega_1^{\text{chain}}$, requires calculating the number of valid configurations where exactly one boundary is up, forcing its immediate neighbor to be down. For a chain of size $N$, fixing one end to $\uparrow\downarrow$ and the opposite end strictly to $\downarrow$ leaves $N-3$ interior sites to be populated by non-adjacent up spins. The number of such configurations is given by the Fibonacci number $F_{N-1}$. Accounting for the two symmetric boundaries, the total macroscopic degeneracy of the first excited state in the open chain is
\begin{align}
\Omega_1^{\text{chain}} = 2 F_{N-1}  
\end{align}

To determine the degeneracy $\Omega_1^{\text{ring}}$, we must place a single $\uparrow\uparrow$ block in a ring of size $N$. To prevent further adjacencies ($N_{\uparrow\uparrow} > 1$), this block must be flanked by down spins, effectively forming a $\downarrow\uparrow\uparrow\downarrow$ sequence that occupies $4$ sites. This sequence can be anchored in $N$ distinct starting positions around the ring. The remaining $N-4$ sites must be filled with non-adjacent up spins, generating $F_{N-2}$ possible arrangements. The total degeneracy for the first excited state of the ring is therefore strictly proportional to the Fibonacci sequence:
\begin{align}
\Omega_1^{\text{ring}} = N F_{N-2} 
\end{align} 

\section{Generalization for chain system}
\subsection{Second-Excited State}\label{subsectionSecond}

Following the energy relations established in Eqs.~\eqref{gap_chain} and \eqref{gap_ring}, the evaluation of the second excited state unveils a systematic difference in the spectral structure of both topologies and exposes the generalized combinatorial pattern for higher-order excitations.

For the open chain, the second excited state corresponds to the next allowable integer solution to the excitation gap $\Delta E^{\text{chain}} = J(2b + 4k)$. Setting $2b + 4k = 4$ yields an energy shift of $\Delta E_2^{\text{chain}} = 4J$. To calculate the exact macroscopic degeneracy, we must account for the spatial distribution of these defects, which effectively partitions the lattice into independent subsystems. This energy level is fundamentally degenerate across two distinct topological classes:
\begin{enumerate}
    \item \textit{Dual boundary excitation} ($b = 2$, $k = 0$): Both boundary spins are pointing up, forcing their immediate neighbors down ($\uparrow\downarrow \dots \downarrow\uparrow$). The available interior space is reduced to $N-4$ sites, all of which are governed by the non-adjacency constraint. The number of microstates for this class is exactly the Fibonacci number  \begin{equation}
        \Omega_{2,(2,0 )}^{\text{chain}} =  F_{N-2} .
    \end{equation}
    \item \textit{Single bulk defect} ($b = 0$, $k = 1$): Both boundaries are anchored downward ($\downarrow$), and exactly one pair of adjacent up spins ($\uparrow\uparrow$) is placed within the interior. To satisfy the constraint, this pair must be isolated by down spins, forming a local $\downarrow\uparrow\uparrow\downarrow$ block. This defect partitions the $N$-site chain into two independent non-adjacent sub-chains: a left sub-chain of length $j$ and a right sub-chain of length $N-2-j$, both of which share the bounding down spins of the defect block. The number of valid configurations for a fixed defect position is the product of their respective Fibonacci degeneracies, $F_{j} F_{N-2-j}$. Summing over all valid interior partitions yields the degeneracy for this class:
\begin{align}
    \Omega_{2,(0,1)}^{\text{chain}} = \sum_{j=1}^{N-3} F_j F_{N-2-j}.
\end{align}
\end{enumerate}

The total degeneracy of the second excited state for the open chain, $\Omega_2^{\text{chain}}$, is the sum of these two combinatorial spaces:
\begin{align}
    \Omega_2^{\text{chain}} = F_{N-2} + \sum_{j=1}^{N-3} F_j F_{N-2-j}. \label{omega2_chain}
\end{align}

\subsection{Third Excited State}
 To determine the combinatorial structure of the third excited state ($\Delta E_3^{\text{chain}} = 6J$), we solve the excitation gap equation $2b + 4k = 6$. Since the number of boundary spins is strictly bounded by $b \in \{0, 1, 2\}$, the equation admits exactly one integer solution:
\begin{align}
    b = 1, \quad k = 1. \label{ecuacion_third_1}
\end{align}
This result dictates that the $6J$ energy level is structurally homogeneous; it consists exclusively of configurations featuring exactly one excited boundary spin and exactly one adjacent up-spin pair within the bulk. The boundary excitation ($b = 1$) can occur at either the left or right edge, yielding a global symmetry factor of $2$. Assuming the left edge is excited, the topological defect ($k = 1$) can either be isolated within the bulk or collapse against the boundary excitation.

If the bulk defect is isolated, the boundary is anchored as $\uparrow_1 \cdots \downarrow_N$, and the interior defect $\uparrow\uparrow\downarrow$ partitions the remaining available space. As derived previously, this isolated partitioning yields a subset of states governed by the convolution $\sum_{j=1}^{N-4} F_j F_{N-3-j}$.
However, if the bulk defect collapses against the excited boundary, it forms a continuous block $\uparrow_1 \uparrow_2 \downarrow_3$. Since only a single adjacent up-spin pair ($\uparrow\uparrow$) is permitted within the chain to satisfy Eq.~\eqref{ecuacion_third_1}, this structural anchoring strictly consumes $3$ sites. This leaves a contiguous segment of length $N-3$ to be tiled by non-adjacent configurations, which explicitly contributes $F_{N-2}$ additional microstates. The total degeneracy is the sum of both classes:
\begin{align}
\Omega_{3,(1,1)}^{\text{chain}} = 2 \left( F_{N-2} + \sum_{j=1}^{N-4} F_j F_{N-3-j} \right). \label{omega3_chain_expanded}
\end{align} 

\subsection{Fourth-excited state}

The combinatorial structure of the fourth excited state ($\Delta E_4^{\text{chain}} = 8J$) requires a strict explicit partitioning of topological clustering, separating boundary collapses from bulk mergers. The topological gap equation $2b + 4k = 8$ admits exactly two macroscopic classes of configurations.

The first class ($b = 2, k = 1$) features excitations at both boundaries ($\uparrow_1$ and $\uparrow_N$) and exactly one adjacent pair within the bulk. The spatial distribution of this single bulk defect splits into three geometrically distinct sub-cases:
\begin{enumerate}
    \item \textit{Left-boundary collapse:} The defect merges with the left boundary, forming $\uparrow_1 \uparrow_2 \downarrow_3$. To prevent further defects, the right boundary segment must end in $\downarrow_{N-1} \uparrow_N$. The remaining free interior length is $N-5$, which yields exactly $F_{N-3}$ non-adjacent configurations.
    \item \textit{Right-boundary collapse:} By symmetry, the defect merges with the right boundary ($\downarrow_{N-2} \uparrow_{N-1} \uparrow_N$), also yielding $F_{N-3}$ distinct configurations.
    \item \textit{Isolated bulk defect:} The defect is fully separated from the boundaries ($\dots \downarrow \uparrow\uparrow \downarrow \dots$). This isolates the defect as a partitioning barrier, generating a first-order discrete convolution of the Fibonacci sequence over an effective volume of $N-4$.
\end{enumerate}
The total degeneracy for the first class is therefore:
\begin{align}
    \Omega_{4,(2,1)}^{\text{chain}} = 2F_{N-3} + \sum_{j=1}^{N-5} F_j F_{N-4-j}. \label{omega4_class1}
\end{align}

The second class ($b = 0, k = 2$) features unexcited boundaries ($\downarrow_1$ and $\downarrow_N$) and exactly two bulk defects. The distribution of these pairs depends on their spatial clustering:
\begin{enumerate}
    \item \textit{Merged triplet:} The two defects collapse into a single continuous block ($\uparrow\uparrow\uparrow$). This single topological barrier partitions the chain into two independent sub-chains over an effective volume of $N-3$, yielding a first-order convolution $\sum_{j=1}^{N-4} F_j F_{N-3-j}$.
    \item \textit{Isolated pairs:} The two pairs are spatially separated ($\cdots\uparrow\uparrow \dots \uparrow\uparrow \cdots$). They act as two independent barriers, partitioning the interior into three distinct sub-chains. This topological arrangement is governed by a second-order convolution over an effective volume of $N-4$.
\end{enumerate}
The total degeneracy for the second class is:
\begin{align}
    \Omega_{4,(0,2)}^{\text{chain}} = \sum_{j=1}^{N-4} F_j F_{N-3-j} + \sum_{i=1}^{N-6} \sum_{j=1}^{N-5-i} F_i F_j F_{N-4-i-j}. \label{omega4_class2}
\end{align}

The exact macroscopic degeneracy of the fourth excited state is the sum of these two partitioned spaces:
\begin{align}
    \Omega_4^{\text{chain}} = 2F_{N-3} &+ \sum_{j=1}^{N-5} F_j F_{N-4-j} + \sum_{j=1}^{N-4} F_j F_{N-3-j} \nonumber \\
    &+ \sum_{i=1}^{N-6} \sum_{j=1}^{N-5-i} F_i F_j F_{N-4-i-j}. \label{omega4_total}
\end{align}
Equation (\ref{omega4_total}) conclusively proves that at higher energies, the critical manifold is constructed by the exact superposition of isolated defects, boundary collapses, and volumetric clustering, strictly adhering to the Fibonacci convolutions defined by the effective partitioned volumes.

\subsection{Fifth-excited state}\label{subsectionThird}

The fifth excited state ($\Delta E_5^{\text{chain}} = 10J$) strictly follows the odd-parity constraint. The gap equation $2b + 4k = 10$ reduces to $b + 2k = 5$, which admits exactly one integer solution: $b = 1$ and $k = 2$. This restricts the macroscopic manifold to configurations with exactly one excited boundary and two bulk defects, introducing a global spatial symmetry factor of $2$.

To determine the exact degeneracy, the spatial clustering of these three topological objects must be strictly partitioned. Assuming the left boundary is excited, four distinct structural sub-cases emerge:
\begin{enumerate}
    \item \textit{Full collapse:} Both bulk defects merge with the excited boundary, forming a continuous block ($\uparrow_1\cdots \uparrow_{i} \uparrow_{i+1} \downarrow_{i+2} \cdots \downarrow_{N}$). This boundary-anchored cluster leaves an unpartitioned effective volume, yielding exactly $F_{N-3}$ microstates.
    \item \textit{Partial boundary collapse:} One defect merges with the boundary, while the second defect remains isolated in the bulk ($ \uparrow_1 \uparrow_2 \downarrow_3 \cdots \uparrow_{i} \uparrow_{i+1} \cdots \downarrow_{N}$). This partitions the interior into two sub-chains, governed by a first-order convolution over an effective volume of $N-4$.
    \item \textit{Isolated triplet:} The boundary is anchored conventionally, and the two bulk defects merge into a single triplet ($ \uparrow_1 \downarrow_2 \cdots \uparrow_{i} \uparrow_{i+1}\uparrow_{i+2} \cdots \downarrow_{N}$) isolated in the interior. This similarly partitions the chain into two sub-chains over an effective volume of $N-5$.
\end{enumerate}

The partial boundary collapse and the isolated triplet mathematically yield the identical convolution sum, allowing them to be grouped. The total macroscopic degeneracy for the fifth excited state is exactly defined by:
\begin{align}
    \Omega_5^{\text{chain}} = 2 \Bigg( F_{N-3} &+ 2\sum_{j=1}^{N-5} F_j F_{N-4-j} \nonumber \\
    &+ \sum_{i=1}^{N-7} \sum_{j=1}^{N-6-i} F_i F_j F_{N-5-i-j} \Bigg). \label{omega5_total}
\end{align} 

 \subsection{Generalization to the $m$-th State of the Open Chain}
\label{sec:generalization}

The explicit constructions detailed in the preceding subsections, from the ground state through the fifth excited state, reveal a unifying combinatorial architecture \cite{flajolet2009analytic}. In this section, we formalize this structure into a closed-form expression for an arbitrary state number $m$. 

\subsubsection{Excitation classes and the Diophantine constraint}

For a macroscopic state at a total excitation energy $\Delta E_m^{\text{chain}} = 2mJ$ above the ground state, the excitation quantum numbers $(b, k)$---where $b$ counts excited boundaries and $k$ counts bulk adjacent pairs---must satisfy the linear Diophantine constraint:
\begin{equation}
    b + 2k = m, \qquad b \in \{0, 1, 2\}, \quad k \in \mathbb{N}_0.
    \label{eq:diophantine}
\end{equation}
The admissible integer solutions depend entirely on the parity of $m$. For even $m$, the solutions are $(b, k) = (0,\, m/2)$ and, provided $m \geq 2$, $(b, k) = (2,\, (m-2)/2)$. For odd $m$, the unique solution is $(b, k) = (1,\, (m-1)/2)$. Each valid solution defines a distinct topological class of configurations that contributes independently to the total macroscopic degeneracy.

\subsubsection{Boundary symmetry factor}

When $b = 1$, the single excited boundary can be located at either the left or the right end of the chain. Because these spatial configurations are related by reflection symmetry but correspond to distinct spin microstates, the $b = 1$ class carries a global symmetry factor $\gamma_1 = 2$. Conversely, classes characterized by $b = 0$ or $b = 2$ are inherently symmetric under reflection, yielding $\gamma_0 = \gamma_2 = 1$.

\subsubsection{Cluster decomposition of bulk defects}

For configurations with $k \geq 1$ bulk adjacent pairs, these defects can merge into $c$ contiguous clusters, where $1 \leq c \leq k$. A topological cluster of $\ell$ adjacent pairs occupies exactly $\ell + 1$ consecutive up spins. Consequently, the number of distinct ways to partition $k$ defects into exactly $c$ ordered clusters is dictated by the standard composition count \cite{flajolet2009analytic}:
\begin{equation}
    \binom{k-1}{c-1}.
    \label{eq:composition}
\end{equation}

These $c$ clusters, along with the $b$ excited boundary sites, effectively partition the chain into $p = c + 1$ independent non-adjacent segments. Each segment of length $j_i$ contributes $F_{j_i}$ valid configurations, according to the fundamental Fibonacci counting established for the ground state in Eq.~\eqref{omega_fibo1}. To preserve the total system size, the lengths of these segments must sum to the effective volume:
\begin{equation}
    V  = N - b - k - c,
    \label{eq:veff}
\end{equation}
where the subtracted terms explicitly account for the fixed spatial footprint of the boundary sites, the excited pair sites, and the cluster-boundary sites.

\subsubsection{The Fibonacci convolution and its boundary extension}

The total number of ways to distribute this effective volume among $p$ segments, with each $j_i \geq 1$, is governed by the $p$-fold Fibonacci convolution:
\begin{equation}
    \mathcal{C}_p(V) = \sum_{\substack{j_1 + \cdots + j_p = V \\ j_i \geq 1}} \prod_{i=1}^{p} F_{j_i}.
    \label{eq:fconv}
\end{equation}
This standard convolution captures the behavior of purely internal defect clusters. However, when a cluster lies adjacent to an excited boundary ($b \geq 1$), the corresponding edge segment can collapse to length zero or even $-1$. Physically, this indicates the spatial merging of the boundary excitation and a bulk cluster, eliminating the non-adjacent segment between them. Utilizing the analytic continuation of the Fibonacci sequence ($F_{-1} = 1$, $F_0 = 0$) \cite{Bunder01051992}, we define the boundary-extended convolution as:
\begin{equation}
    \widetilde{\mathcal{C}}_p^{(b)}(V) = \sum_{\substack{j_1 + \cdots + j_p = V \\ j_i^{(\mathrm{edge})} \geq -1,\; j_i^{(\mathrm{int})} \geq 1}} \prod_{i=1}^{p} F_{j_i},
    \label{eq:fconv_ext}
\end{equation}
where $j_i^{(\mathrm{edge})}$ denotes the segments adjacent to excited boundaries (the first segment for $b \geq 1$, and additionally the last segment for $b = 2$), while $j_i^{(\mathrm{int})}$ denotes all strictly interior segments. This yields the explicit forms:
\begin{align}
    \widetilde{\mathcal{C}}_p^{(0)}(V) &= \mathcal{C}_p(V), \\[4pt]
    \widetilde{\mathcal{C}}_p^{(1)}(V) &= \sum_{j_1=-1}^{V-p+1} F_{j_1}\, \mathcal{C}_{p-1}(V - j_1), \quad p \geq 2, \\[4pt]
    \widetilde{\mathcal{C}}_p^{(2)}(V) &= \sum_{j_1=-1}^{V-p+3}\; \sum_{j_p=-1}^{V-j_1-p+2} F_{j_1}\, F_{j_p}     \nonumber \\
    &\quad \times \mathcal{C}_{p-2}(V - j_1 - j_p), \quad p \geq 3.
    \label{eq:fconv_b2}
\end{align}

A critical subtlety arises regarding the summation limits in Eq.~\eqref{eq:fconv_b2}. When $b = 2$, both edge segments are free to take values greater than or equal to $-1$. To ascertain the rigorous upper bound for $j_1$, all other segments must be minimized: the $p - 2$ interior segments contribute a minimum value of $1$, while the final edge segment $j_p$ contributes a minimum of $-1$. The geometric constraint $j_1 + j_p + \sum_{i=2}^{p-1} j_i = V$ then implies:
\begin{equation}
    j_1 \leq V - (p-2)(1) - (-1) = V - p + 3.
    \label{eq:critical_bound}
\end{equation}
This extended limit accounts for the final edge segment absorbing one additional spatial unit by collapsing to $j_p = -1$. Neglecting this correction systematically undercounts configurations where both boundary clusters fully merge with their respective excited boundaries, an effect that becomes dominant for higher excitations ($m \geq 6$).
 
 \subsubsection{General formula and physical bounds}

 By combining the topological cluster decomposition with the boundary-extended convolutions, the degeneracy corresponding to a specific class $(b, k)$ is given by:
\begin{equation}
    D(N, b, k) =
    \begin{cases}
        F_{N-b}, & k = 0, \\[6pt]
        \displaystyle\sum_{c=1}^{k} \binom{k-1}{c-1}\, \widetilde{\mathcal{C}}_{c+1}^{(b)}(N - b - k - c), & k \geq 1.
    \end{cases}
    \label{eq:class_deg}
\end{equation}

Another constraint is that the physical upper bound of this combinatorial framework is dictated by the finite dimension of the Hilbert space. The excitation spectrum comprises exactly $2N-1$ distinct energy levels, bounding the excitation index to $0 \leq m \leq 2N $. To satisfy the completeness relation $\sum_{m=0}^{2N-2} \Omega_m^{\mathrm{chain}}(N) = 2^N$, the macroscopic degeneracy condenses into a compact formula
\begin{align}
    \Omega_m^{\mathrm{chain}}(N) &=
        \displaystyle \sum_{\substack{b+2k = m \\ b \in \{0,1,2\}, k \geq 0}} \gamma_b \, D(N, b, k), & 0 \leq m \leq 2N. 
      \label{eq:omega_general}
\end{align}
with symmetry factors $\gamma_0 = \gamma_2 = 1$ and $\gamma_1 = 2$.
A non-trivial consequence of this formulation is that the antepenultimate and penultimate excitation levels of the open chain exhibit exactly zero degeneracy ($\Omega_{2N-2}^{\mathrm{chain}} = \Omega_{2N-1}^{\mathrm{chain}} = 0$). This physical behavior stems directly from the topological constraints of the lattice. The general pattern dictates that the chain spectrum contains exactly $2N-1$ accessible macroscopic levels distributed across the indices $m \in \{0, 1, \dots, 2N-3\} \cup \{2N\}$, resulting in a strict spectral gap spanning $m = 2N-2$ and $m = 2N-1$. This exclusion is physically intuitive: the fully polarized state (all spins pointing up) inherently possesses $b=2$ boundary excitations and $k=N-1$ bulk adjacent pairs. According to the Diophantine relation, this yields an absolute maximum excitation index of $m = b + 2k = 2 + 2(N-1) = 2N$. It is topologically impossible to arrange excitations to achieve exactly $m = 2N-2$ or $m = 2N-1$, as introducing a single down-spin into the fully polarized state necessarily breaks multiple adjacent bonds simultaneously, dropping the index directly to a maximum of $m = 2N-3$. Consequently, this establishes the fundamental property that $\Omega_{2N}^{\mathrm{chain}}(N) = 1$, corresponding to the unique configuration where all spins align upwards.

This unified formulation successfully reproduces the derived low-energy states, providing systematic consistency checks. For the ground state ($m = 0$), the sole class is $(0, 0)$, cleanly returning $\Omega_0 = F_N$. For the first excited state ($m = 1$), the unique class $(1, 0)$ correctly recovers $\Omega_1 = 2 F_{N-1}$. For $m = 2$, evaluating the classes $(0, 1)$ and $(2, 0)$ precisely reproduces $\Omega_2 = \widetilde{\mathcal{C}}_2^{(0)}(N-1) + F_{N-2}$. Analogous reductions rigorously recover all explicit combinatorial forms derived for $m \in \{3, 4, 5\}$. 

Nevertheless, the explicit evaluation of these nested convolutions $\widetilde{\mathcal{C}}_p^{(b)}$ for large $m$ becomes combinatorially complex, effectively motivating the algebraic transfer-matrix approach developed in the subsequent section.

  \section{Exact Algebraic Generalization via Transfer Matrix}
\label{sec:transfer}

\subsection{Construction of the transfer matrix}

While the explicit summation of topological clusters provides deep physical insight into the fractionalization of the critical manifold, expanding the nested discrete convolutions for an arbitrary $m$-th excited state becomes combinatorially intractable. To achieve a strictly programmable and universally exact generalization, we map the one-dimensional topological defect onto an algebraic generating function using the transfer matrix formalism \cite{baxter2007exactly,flajolet2009analytic}.

The energy gap relative to the ground state at $B/J=2$ is governed by the local topological score $m = b + 2k$. We introduce a formal generating-function variable $y$ to track these excitation units. In the basis $\{|\!\downarrow\rangle, |\!\uparrow\rangle\}$, the transfer matrix $\mathbf{T}$ encodes the local Boltzmann weights between adjacent sites $i$ and $i+1$:
\begin{equation}
    \mathbf{T} = \begin{pmatrix} 1 & 1 \\ 1 & y^2 \end{pmatrix}.
    \label{eq:transfer_matrix}
\end{equation}

Here, the entries $\mathbf{T}_{\downarrow\uparrow} = \mathbf{T}_{\uparrow\downarrow} = \mathbf{T}_{\downarrow\downarrow} = 1$ reflect that these pairs create no additional bulk frustration. Conversely, $\mathbf{T}_{\uparrow\uparrow} = y^2$ accounts for the two excitation units contributed by a bulk adjacent pair.

Fractional excitations at the boundaries must be treated independently. An up-spin at either boundary contributes exactly $1$ unit to the excitation score, carrying a weight of $y$. Down-spins at the boundaries carry a weight of $1$. This establishes the left and right boundary state vectors as $\mathbf{v}_L = \begin{pmatrix} 1 & y \end{pmatrix}$ and $\mathbf{v}_R = \begin{pmatrix} 1 & y \end{pmatrix}^T$.

The canonical generating polynomial for an open chain of $N$ sites is the exact contraction of the boundary vectors with the $(N-1)$-th power of the transfer matrix:
\begin{equation}
    Z_N^{\mathrm{chain}}(y) = \mathbf{v}_L \mathbf{T}^{N-1} \mathbf{v}_R.
    \label{eq:chain_genfun}
\end{equation}
The macroscopic degeneracy of the $m$-th excited state is then mathematically extracted as the exact coefficient of the $y^m$ term:
\begin{equation}
    \Omega_m^{\mathrm{chain}}(N) =  
        \left[ y^m \right] Z_N^{\mathrm{chain}}(y).
    \label{eq:omega_general_tm}
\end{equation}

\subsection{Diagonalization and closed-form generating functions}

To extract closed-form analytical expressions, the matrix power is resolved via diagonalization, $\mathbf{T} = \mathbf{P} \mathbf{\Lambda} \mathbf{P}^{-1}$. The eigenvalues are determined by the characteristic equation $\det(\mathbf{T} - \lambda \mathbf{I}) = 0$, which yields:
\begin{equation}
    \lambda_{\pm} = \frac{1 + y^2 \pm \Delta(y)}{2}, \qquad \Delta(y) \equiv \sqrt{y^4 - 2y^2 + 5}.
    \label{eq:eigenvalues}
\end{equation}
The corresponding right eigenvectors are $\mathbf{v}_{\pm} = (1, \lambda_{\pm} - 1)^T$. The diagonalization renders the matrix power trivial ($\mathbf{T}^{N-1} = \mathbf{P} \mathbf{\Lambda}^{N-1} \mathbf{P}^{-1}$). Substituting this back into Eq.~\eqref{eq:chain_genfun}, the chain generating function becomes a sum of two geometric terms:
\begin{equation}
    Z_N^{\mathrm{chain}}(y) = A_+(y)\, \lambda_+^{N-1} + A_-(y)\, \lambda_-^{N-1},
    \label{eq:chain_spectral}
\end{equation}
where the boundary-dependent amplitudes are given by $A_{\pm}(y) = (\mathbf{v}_L \cdot \mathbf{v}_{\pm})\,(\mathbf{v}_{\pm}^{-1} \cdot \mathbf{v}_R)$, with $\mathbf{v}_{\pm}^{-1}$ denoting the corresponding row of $\mathbf{P}^{-1}$. Substituting this closed-form expression into Eq.~\eqref{eq:omega_general_tm} yields the exact analytical degeneracies of the open chain.

For the closed-ring topology with periodic boundary conditions, the generating function is simply the trace of the transfer matrix:
\begin{equation}
    Z_N^{\mathrm{ring}}(y) = \mathrm{Tr}(\mathbf{T}^N) = \lambda_+^N + \lambda_-^N,
    \label{eq:ring_genfun}
\end{equation}
and the corresponding macroscopic degeneracy is extracted as  
\begin{equation}
\Omega_k^{\mathrm{ring}}(N)  =  
        [y^{2k}]\,(\lambda_+^N + \lambda_-^N) .
    \label{eq:omega_general_ring_tm}
\end{equation}
Ultimately, this algebraic formulation intrinsically absorbs all discrete Fibonacci convolutions, automatically computing every valid spatial partition without requiring piecewise cases. For arbitrary polynomial extraction, this method resolves the entire critical spectrum in logarithmic computational time, proving the exact isomorphism between the macroscopic critical degeneracies and the generating polynomial of local fractional defects. Furthermore, leveraging these exact analytical degeneracies and energies, the specific heat and entropy for both topologies are plotted in the Appendix. As observed, both systems exhibit a zero-temperature entropy proportional to $k_B \ln [\Omega_0(N)]  $, governed by the Fibonacci sequence for the open chain and the Lucas sequence for the closed ring.

 \section{Exact Degeneracies for the Ring Topology}
\label{app:ring}

The combinatorial analysis for the one-dimensional Ising ring fundamentally differs from the open chain due to periodic boundary conditions ($\sigma_{N+1} \equiv \sigma_1$), which inherently eliminate fractional boundary excitations. Consequently, the excitation spectrum is governed exclusively by bulk adjacent pairs incurring an energy cost of $\Delta E = 4J$.

For the $m$-th excited state, the $m$ bulk defects organize into $c$ contiguous clusters, partitioning the ring into exactly $c$ antiferromagnetic arcs over an effective volume $V_{\mathrm{eff}} = N - m - c$. Factoring in the composition of these clusters, $\binom{m-1}{c-1}$, and the circular rotational symmetry, $N/c$, the exact macroscopic degeneracy condenses into a single closed-form piecewise expression. Similar to the open chain, the finite dimension of the Hilbert space strictly bounds the spectrum, isolating the absolute maximum excitation of the fully polarized ring ($m=N$) as a unique configuration:
\begin{equation}
    \Omega_m^{\mathrm{ring}}(N) =
    \begin{cases}
        L_N, & m = 0, \\[8pt]
        \displaystyle\sum_{c=1}^{m} \binom{m-1}{c-1}\, \frac{N}{c}\, \mathcal{C}_c(N - m - c), & 1 \leq m < N , \\[8pt]
        1, & m=N
    \end{cases}
    \label{eq:omega_ring}
\end{equation}
This structural formulation demonstrates that the absence of boundary defects rigorously simplifies the partition sum. It eliminates the need for boundary-extended convolutions ($\widetilde{\mathcal{C}}_p^{(b)}$), ensuring the convolution order precisely matches the cluster count. Explicit low-energy cases naturally emerge from this expression---such as $\Omega_1^{\mathrm{ring}}(N) = N F_{N-2}$---and strictly agree with the algebraic trace evaluated via the transfer matrix formalism, $\Omega_m^{\mathrm{ring}}(N) = [y^{2m}]\,\mathrm{Tr}(\mathbf{T}^N)$.

A consequence derived from Eq.~\eqref{eq:omega_ring} is that the penultimate excitation level of the closed ring exhibits exactly zero degeneracy ($\Omega_{N-1}^{\mathrm{ring}} = 0$). Physically, this implies that the system is strictly forbidden from occupying this specific state, resulting in a non-trivial spectral gap of $8J$ just below the highest accessible energy level. This topological restriction arises because achieving the penultimate excitation requires exactly $k = N_{\uparrow\uparrow} = N-1$ adjacent up-spin pairs. However, introducing a single down-spin into the fully polarized ferromagnetic state ($k=N$) necessarily breaks two adjacent bonds, dropping the number of pairs directly to $k = N-2$. As a result, the geometric constraint of the closed ring makes the condition $k = N-1$ topologically impossible. Therefore, no physical microstate exists at this energy level, an exclusion strictly predicted by the combinatorial formula yielding zero degeneracy. Ultimately, this establishes the fundamental property that $\Omega_{N}^{\mathrm{ring}}(N) = 1$, corresponding to the unique configuration where all spins align upwards.

A comprehensive numerical comparison demonstrating the exact agreement between the analytical formulas in Eqs.~\eqref{eq:omega_general} and \eqref{eq:omega_ring}, the transfer-matrix solutions in Eqs.~\eqref{eq:omega_general_tm} and \eqref{eq:omega_general_ring_tm}, and the exact degeneracies obtained through computational simulations is provided in the Supplemental Material~\cite{supp}. Furthermore, as a consequence of these differing macroscopic degeneracies, we conducted a comparative thermodynamic analysis of the specific heat and entropy (see Appendix~\ref{appendix_B}). Notably, the distinct ground-state degeneracies inherent to each topology manifest as clear differences in the zero-temperature residual entropy. Such pronounced macroscopic degeneracies hold significant potential for the design of discrete quantum heat machines, as the entropic shifts associated with traversing these critical manifolds can be strategically leveraged to substantially enhance heat jumps. Additionally, the source codes used to generate the Supplemental Material and the corresponding thermodynamic plots are openly accessible, as referenced in the Data Availability section.

 \section{Conclusion}
\label{sec:conclusion}
In this manuscript, we have established exact, closed-form combinatorial expressions governing the complete excitation spectrum of the one-dimensional antiferromagnetic Ising model at the level-crossing (LC) point $B/J = 2$. By systematically analyzing topological frustration and non-adjacency constraints, we demonstrated that the macroscopic degeneracies of both open chains and periodic rings share a unifying mathematical architecture.

Specifically, the density of states is constructed through integer compositions of topological defects weighted by Fibonacci convolutions. The closed ring strictly follows standard convolutions rooted in the Lucas sequence, whereas the open chain incorporates boundary-extended convolutions to account for fractional edge excitations. Both structural frameworks have been independently verified through algebraic extraction via the transfer-matrix formalism and exact computational enumeration.

Furthermore, this rigorous state counting exposes non-trivial spectral gaps near the fully polarized limit. Dictated by the topological constraints of breaking adjacent bonds, our analytical formulas strictly predict vanishing degeneracies for the two penultimate macroscopic energy levels in the open chain and the single penultimate level in the closed ring, physically forbidding the system from occupying these specific states.

Ultimately, these results reveal that the entire critical manifold inherits the recursive Fibonacci structure of the ground state. The derived exact formulas bypass combinatorial intractability, providing a rigorous analytical foundation for exploring finite-size scaling, residual entropy, and the exact thermodynamic limits of discrete spin systems at quantum criticality. By analytically reconstructing the density of states, the exact residual critical entropy for every accessible energy level can be directly obtained. Future work should aim to investigate the excited states associated with even degeneracies and explore the extension of this combinatorial formalism to other spin models and lattice topologies.

\begin{acknowledgments}
{Authors acknowledge financial support from ANID
Fondecyt grant no. 1240582 and 1250173}. B.C. acknowledges PUCV. B.C and M.H.G acknowledge ''Direcci\'on de Postgrado'' of UTFSM. {B.C. acknowledges the support of ANID Becas/Doctorado Nacional 21250015}. Authors acknowledge partial support from CEDENNA CIA 250002.
\end{acknowledgments}

\section*{DATA AVAILABILITY}
The data that support the findings of this article are openly available \cite{castorene_2026_19741173}, embargo periods may apply.

\appendix
\section{Parity-Dependent Ground Microstates} \label{parity}

A secondary Level Crossing emerges at $B_{\mathrm{crit}} = B/J = 1$. Minimizing the Hamiltonian in this regime requires maximizing the domain walls $N_{\uparrow\downarrow}$ relative to $N_\uparrow$, strictly enforcing the non-adjacency constraint $N_{\uparrow\uparrow} = 0$. Consequently, the ground-state degeneracy is dictated entirely by finite-size parity and boundary topology.

For the open chain, an odd $N$ perfectly accommodates a single alternating configuration anchored by boundary down-spins ($\downarrow\uparrow\downarrow\dots\uparrow\downarrow$), yielding no macroscopic degeneracy ($\Omega_{\text{ground}}^{\text{chain}} = 1$). For an even $N$, the ground manifold comprises the two perfectly alternating Néel states plus $N/2$ configurations where non-adjacent interior excitations are bounded by down-spins. This results in a strict linear scaling:
\begin{equation}
    \Omega_{\text{ground}}^{\text{chain}} (N \text{ even}) = \frac{N}{2} + 2. \label{omega_even_chain}
\end{equation}

For the closed ring, periodic boundary conditions naturally yield $N_{\uparrow\downarrow} = 2N_\uparrow$. An even $N$ permits perfect antiferromagnetic alternation, resulting strictly in the two trivial Néel states ($\Omega_{\text{ground}}^{\text{ring}} = 2$). However, an odd $N$ introduces geometric frustration. 

The non-adjacency constraint limits the system to configurations containing exactly one pair of adjacent down spins ($\downarrow\downarrow$). Evaluating the circular combinatorial arrangement for $k = (N-1)/2$ non-adjacent up spins yields:
\begin{equation}
    \Omega_{\text{ground}}^{\text{ring}} (N \text{ odd}) = \frac{N}{N - \frac{N-1}{2}} \binom{N - \frac{N-1}{2}}{\frac{N-1}{2}} = N. \label{omega_odd_ring}
\end{equation}

Ultimately, these relations demonstrate that at $B/J=1$, topological frustration and parity dictate a purely linear scaling mechanism, breaking the recursive Fibonacci-Lucas complexity observed at the primary LC ($B_{\mathrm{crit}}=2$).

\section{Heat capacity and Entropy} \label{appendix_B}

\begin{figure}[H]
    \centering
    \includegraphics[width=0.9\linewidth]{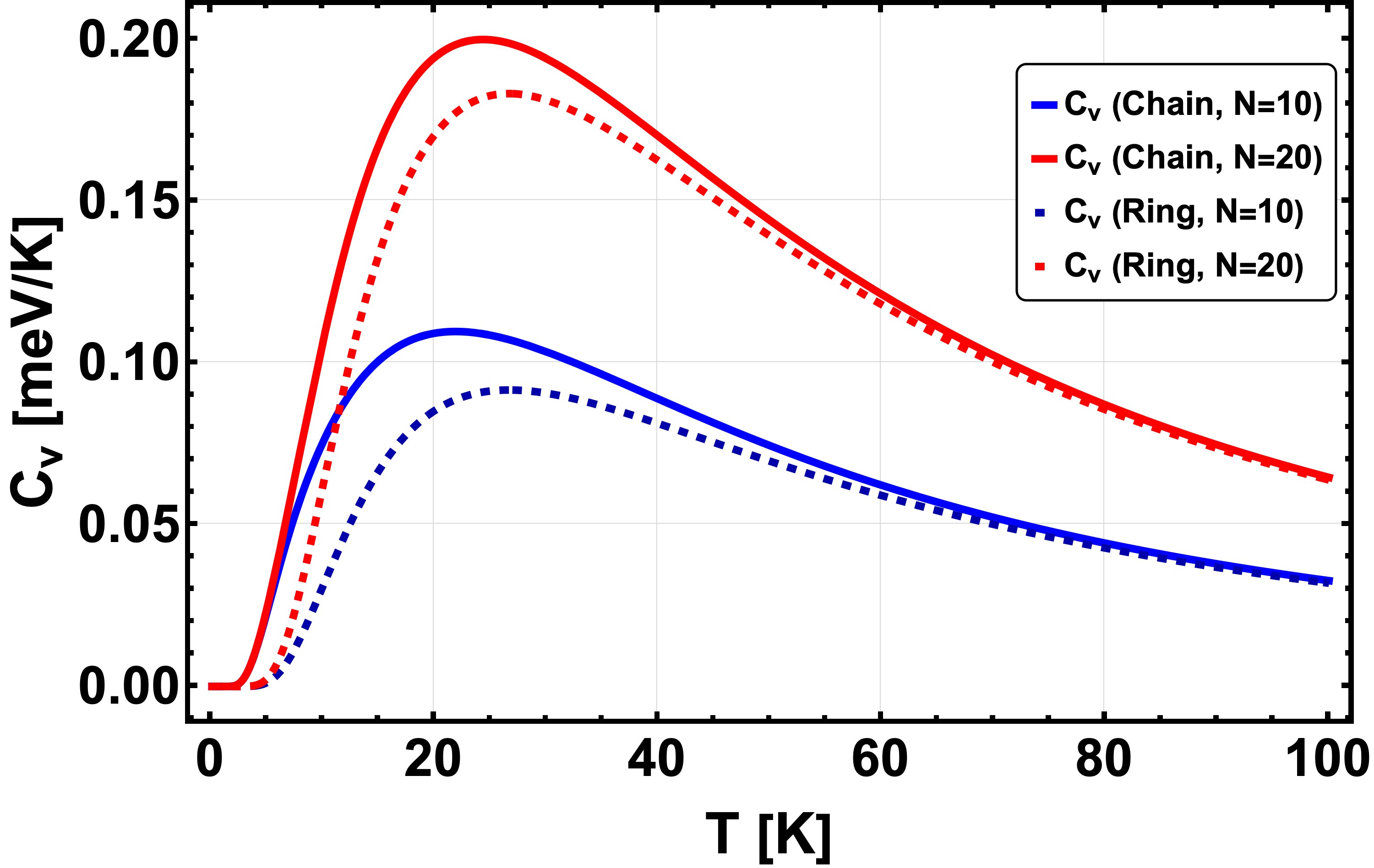}
    \caption{Heat capacity as a function of temperature for the chain and ring systems for different lengths $N$ in units of exchange $J$. The thick (dashed) lines represent the chain (ring) topology for $N=10$ and $N=20$, respectively.}
    \label{fig:HeatCapacity}
\end{figure}

\begin{figure}[H]
    \centering
    \includegraphics[width=0.9\linewidth]{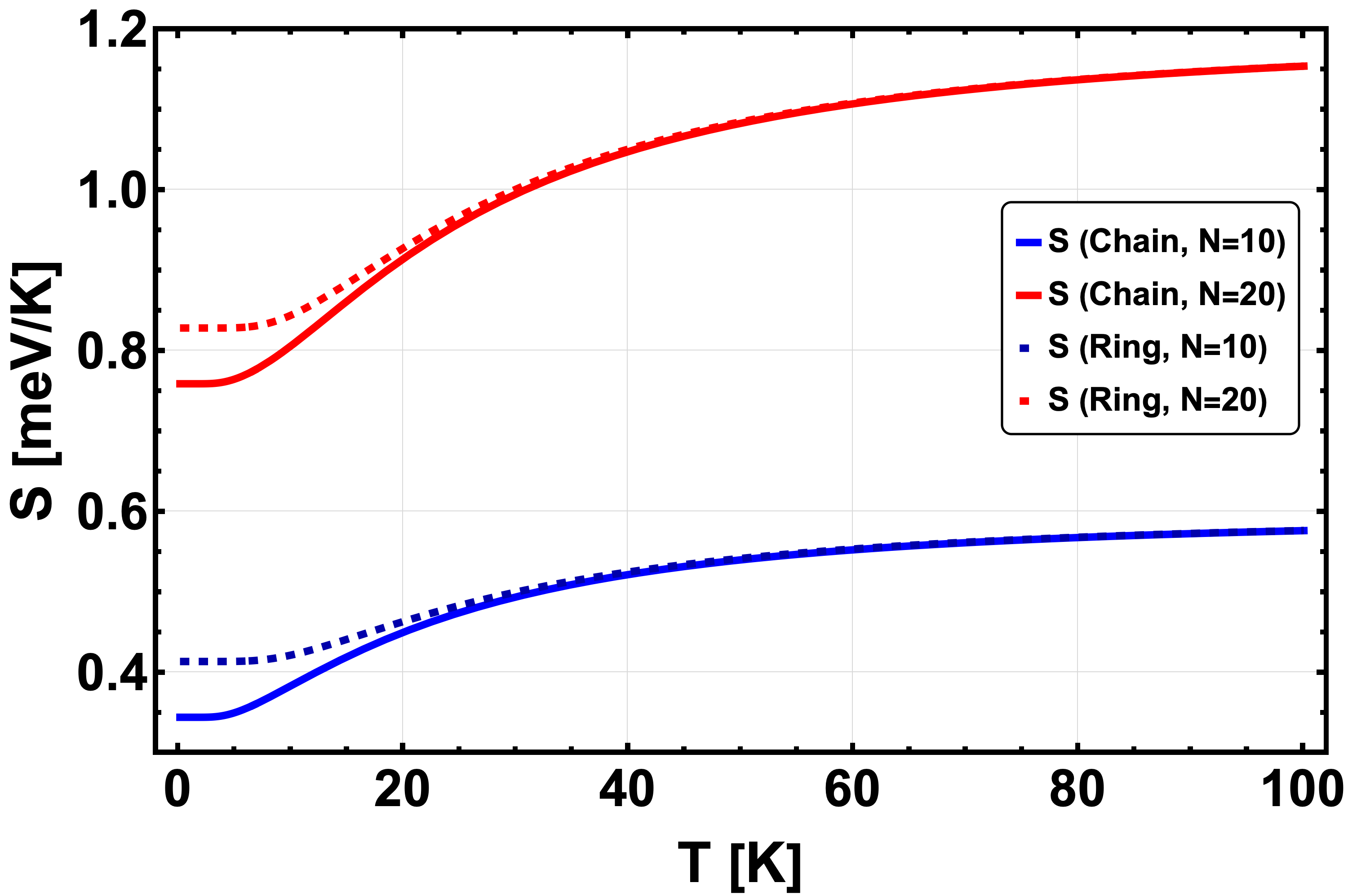}
    \caption{Entropy as a function of temperature for the chain and ring systems for different lengths $N$ in units of exchange $J$. The thick (dashed) lines represent the chain (ring) topology for $N=10$ and $N=20$, respectively.}
    \label{fig:Entropy}
\end{figure}

Using the macroscopic degeneracies for the chain and ring topologies given in Eqs.~\eqref{eq:omega_general} and \eqref{eq:omega_ring}, the exact canonical partition function can be computed as
\begin{align}
    Z(T,N) &= \sum_{m} \Omega_m(N) e^{-\frac{E_m}{k_B T}}.
\end{align}
From the partition function, the exact thermodynamic quantities, such as the heat capacity $C_v(T,N)$ and the entropy $S(T,N)$, are derived using the standard relations:
\begin{align}
    C_v(T,N) &= \frac{\partial}{\partial T} \left( k_B T^2 \frac{\partial \ln Z}{\partial T} \right)  , \\
    S(T,N) &= k_B \ln Z + k_B T \frac{\partial \ln Z}{\partial T}.
\end{align}
These thermodynamic quantities are shown in Figs.~\ref{fig:HeatCapacity} and \ref{fig:Entropy}.

As shown in Fig.~\ref{fig:HeatCapacity}, the heat capacity of all systems starts at zero. At intermediate temperatures, proximate maxima emerge for both topologies, with values that vary with system length. At high temperatures, however, the heat capacities of all systems converge to the same value. In contrast, the entropy shown in Fig.~\ref{fig:Entropy} demonstrates that both systems exhibit a zero-temperature entropy proportional to $k_B \ln [\Omega_0(N)]$, governed by the Fibonacci and Lucas sequences for the open chain and closed ring, respectively. As the temperature increases, only systems of the same length converge to identical asymptotic values, in contrast to the uniform high-temperature convergence observed for the heat capacity.

\bibliography{biblio_3qubit.bib}

\clearpage
\onecolumngrid

\end{document}